\documentclass[pre,aps,twocolumn,showpacs]{revtex4}
\usepackage{graphicx}
\usepackage{amsmath}
\usepackage{amssymb}
\def\slaninafigdir{.}
\begin{document}
\title{%
Equivalence of replica and cavity methods for computing spectra of sparse random matrices
}
\author{%
Franti\v{s}ek Slanina
}
\affiliation{%
Institute of Physics,
 Academy of Sciences of the Czech Republic,
 Na~Slovance~2, CZ-18221~Praha,
Czech Republic
}
\email{
slanina@fzu.cz
}
\begin{abstract}
We show by direct calculation that the replica and cavity methods are
exactly equivalent for the spectrum of Erd\H{o}s-R\'enyi random
graph. We introduce a variational formulation based on the cavity
method and use it to find approximate solutions for the density of
eigenvalues. We also use this variational method for calculating spectra of sparse
covariance matrices.
\end{abstract}
\date{\today}%
\maketitle%
\section{Introduction}
Random matrix theory is a discipline with wide range of physical
applications and plenty of beautiful mathematical  results \cite{mehta_91}. One of the
aspects which makes the problem extremely complex is the fact that
real  physical systems are embedded in three-dimensional Euclidean
space. Their Hamiltonian is often a random matrix, but the randomness
is constrained in a highly non-trivial way. 

The constraints are relatively less severe
in the atomic nucleus, where the
three-dimensionality of physical space is of secondary
importance. Hence the spectacular success of the early works in random
matrix theory, due to Wigner \cite{wigner_55,wigner_57} and Dyson
\cite{dyson_62}. On the other hand, the fundamental constraint arising
from two-body character of the interaction within the (model of an)
atomic nucleus induces several drastic changes
\cite{fre_won_70,bog_flo_71,bog_flo_71a,won_fre_72,bog_flo_fre_gia_mel_won_74}. Most
importantly, the density of states is not a semi-circle, as suggested
by  Wigner, but rather it follows a Gaussian shape. Therefore, sharp
band edges are missing and Lifschitz tails develop. For the current
state of the problem, see e. g. the review \cite{wei_mit_09}.

Even more complicated situation arises in 
 all
random extended systems, like disordered or amorphous semiconductors,
where we  must
take into account the Euclidean constraints. 
Perhaps the easiest of these constraints is the sparsity of the
Hamiltonian matrix, which is due to finite range of interactions. If
we forget the even more severe complications due to precise number
of spatial dimensions (in reality one, two or three), we are left with
the problem of determining the spectrum of a random sparse matrix.

Important breakthrough was achieved using the 
 replica method, which was introduced in the context of random matrices
in \cite{edw_jon_76}. Rodgers and Bray, in their classical work
\cite{rod_bra_88}, solved the problem in the sense that they found an
integral equation for a quantity from which the density of states is
readily obtained. Unfortunately, that equation still resists all
attempts for exact analytic solution. In \cite{rod_bra_88},
two approximative solutions were found. First, in the form of a series
expansion, whose leading term coincides with the Wigner semicircle
law. Second, using a non-perturbative argument, introduced earlier in
\cite{kim_har_85}, the shape of the Lifschitz 
tails in the density of states was found.

The replica method for treating spectra of 
sparse matrices was further developed 
\cite{bra_rod_88,mez_par_zee_99,cav_gia_par_99,cav_gia_par_00,monasson_99,bi_mo_99,sem_cug_02,dean_02,fortin_05,rod_aus_kah_kim_05,cic_orl_06,nag_tan_07,nag_rod_08,kuhn_08,bianconi_08}. Especially, 
the variational formulation of the replica equations
\cite{bi_mo_99,sem_cug_02,kuhn_08} enabled generating self-consistent
approximations, namely the effective medium approximation (EMA), which
is analogous to the coherent potential approximation used for
electrons in random potential. In these approximations, Lifschitz
tails in the spectrum are absent. Further sophistication of the method
consists in the single defect approximation (SDA), which obtains the
Lifschitz tail in the form of infinite sequence of delta-peaks. 

The complexity of the problem becomes evident when we compare these
results with the density of states obtained by numerical
diagonalization of large sample matrices
\cite{evangelou_92,eva_eco_92,bau_gol_01,golinelli_03,rod_cas_kuh_tak_08,kuhn_08}. The
Lifschitz tail is smooth, while the bulk of the density of states is
the combination of continuous component with a set of delta-peaks. The
most marked of these peaks is at the origin, others at eigenvalues
$z=\pm 1$, $\pm\sqrt{2}$ etc. All these structures should emerge from
the solution of the Rodgers-Bray integral equation, but EMA, as well
as SDA, miss all of them. The set of delta-peaks was studied separately
in \cite{sem_cug_02,golinelli_03}, but a theory which would combine naturally
both these peaks and the continuous component is still unavailable.

More recently, spectra of sparse matrices encoding the structure
of random graphs were studied successfully using the cavity approach
(see e. g. \cite{ciz_bou_94}). 
It is based on the fact that large random graphs are locally
isomorphic to trees. This was used e. g. in
\cite{dor_gol_men_sam_03,dor_gol_men_sam_04,sam_dor_men_08} to
calculate spectra of adjacency matrix and Laplacian on complex
networks. In \cite{dor_gol_men_sam_03,dor_gol_men_sam_04}, a
``self-consistent'' version of SDA was used to obtain asymptotic shape
of the Lifschitz tails, which decay as a power law in the case of
scale-free networks. In \cite{sam_dor_men_08} a more sophisticated
calculation lead to an integral equation similar to Rodgers and Bray's
\cite{rod_bra_88}, from which the asymptotics of Lifschitz tails is
found. The cavity method provides an easy way
\cite{rod_cas_kuh_tak_08} 
to obtain the
Wigner semicircle law, as well as the Mar\v{c}enko-Pastur law for
spectrum of covariance matrices. 
It can be also used as an efficient numerical procedure
\cite{rod_cas_kuh_tak_08}, reproducing all peculiarities of the
density of states, including Lifschitz tails and delta-peaks.
 The mathematical justification for the use of the cavity
approach can be found in \cite{bor_lel_08}.

Very powerful method for computing spectral properties of random
matrices is based on supersymmetry and was developed in
\cite{efetov_83,ver_wei_zir_85} (see also the review
\cite{guh_gro_wei_98} and a recent development in
\cite{bun_efe_kra_yev_zir_07}).
 Initially, the results of replica and
supersymmetric methods were found in conflict, which resulted in
serious criticism of the replica trick in general
\cite{ver_zir_85}. Density of states of sparse random matrices was
calculated using supersymmetry \cite{rod_ded_90}, leading to an
equation which was later \cite{fyo_mir_91} shown equivalent to the
replica result of \cite{rod_bra_88}. However, the correlation of
eigenvalues, which was investigated in \cite{mir_fyo_91} using
supersymmetry for the case of
sparse matrices, was not reproduced correctly in replica method, until
the integral over all saddle points was properly taken in
\cite{kam_mez_99}. Since then, the replica method regained its
reputation as an equivalent alternative to supersymmetric
methods. This was further supported by a series of papers 
\cite{kanzieper_01,kanzieper_09,osi_kan_07}.
Finally, let us only mention the works which approach the density of
states by computing  exactly the moments
\cite{kho_rod_97,bau_gol_01}.

In this paper, we show an alternative method to obtain the
Rodgers-Bray integral equation using cavity approach. Therefore, we
prove exact equivalence of replica and cavity method in this
case, which was previously assumed only on the basis of topological
considerations concerning random Erd\H{o}s-R\'enyi graphs. Moreover,
as an important by-product of this proof, we present a variational
formulation of the problem, which serves as useful generator of
self-consistent approximations.

\section{Projector method}

We shall investigate the spectrum of the adjacency matrix $L$ of an Erd\H
os-R\'enyi random graph with $N$ vertices. Therefore, the probability distribution
of the matrix elements factorizes
\begin{equation}
\pi(L)=\prod_{i<j}\Big[\pi_1(L_{ij})\delta(L_{ij}-L_{ji})\Big]\prod_{i}\delta(L_{ii})
\label{eq:distr-matrix}
\end{equation}
where the probability density for a single off-diagonal element is
\begin{equation}
\pi_1(x)=\Big(1-\frac{\mu}{N}\Big)\delta(x)+\frac{\mu}{N}\delta(x-1)\;.
\label{eq:distr-matrixelement}
\end{equation}
The key ingredient of all subsequent analysis is the resolvent
\begin{equation}
R(z)=(z-L)^{-1}
\end{equation}
and its average $\langle{R}(z)\rangle$ over disorder, 
taken with the distribution (\ref{eq:distr-matrix}). It
contains information on the average density of states (here we assume $z$ on the real axis)
\begin{equation}
\mathcal{D}(z)=\lim_{\epsilon\to0^+}\frac{1}{N\pi}\mathrm{Tr}\,\langle{R}(z-i\epsilon)\rangle\;.
\end{equation}

In the spirit of the cavity method, we focus on a single vertex,
surrounded by the rest of the graph.  To calculate the diagonal element of the resolvent on this
vertex, we use the  projector method, formulated generally in  \cite{lowdin_62}. For a different
route which also leads to equivalent results, see \cite{ciz_bou_94}.
Let us have an arbitrary  
projector $P$ and its complement ${P}^C\equiv
1-P$. Then, the projected resolvent is \cite{lowdin_62}
\begin{equation}
PRP=\frac{P}{P(z-L)P-PL{P}^C\,\frac{{P}^C}{z-L} \,{P}^CLP}\;.
\label{eq:partitioning-theorem}
\end{equation}

We denote the singled-out vertex
as $i=0$. Let $P_0$ be the projector to this  vertex.
Furthermore, denote $i=1,2,\ldots,k$  neighbors of the vertex $0$ on the graph
 represented by the matrix $L$ and denote
also $P_{i}$ projector to the neighbor $i$. Let us use
composite indices for other vertices. If  $k_i$ is the number of
neighbors of $i$, denote $[i,1],[i,2],\ldots,[i,k_i-1]$
the neighbors of vertex $i$, except the vertex $0$. The projectors to
the second neighbors of $0$ will be denoted using these indices, so
$P_{i,i'}$ is projector on the vertex $[i,i']$. By analogy, we
define the projectors to third, fourth etc. neighbors of $0$. Note
that on a general graph, some of the projectors may coincide, due to
the presence of cycles.

The cavity approach consists in replacing the graph by a tree, which
is locally isomorphic to it, i. e. neglecting all cycles on the graph.
 Algebraically, it is equivalent to the assumption that the
 complementary projectors can be written as direct sums of projectors
 corresponding to separate branches of the tree  
\begin{equation}
\begin{split}
P_0^C\,&=P_{(1)}\oplus P_{(2)}\oplus\ldots\oplus P_{(k)}\\
P_{(i)}P_i^C\,&=P_{(i,1)}\oplus P_{(i,2)}\oplus\ldots\oplus P_{(i,k_i-1)}\\
P_{(i,i')}P_{i,i'}^C\,&=P_{(i,i',1)}\oplus P_{(i,i',2)}\oplus\ldots\oplus P_{(i,i',k_{i,i'}-1)}\\
&\vdots
\end{split}
\label{eq:assume-tree}
\end{equation}
where $P_{(i)}P_i=P_i$, $P_{(i,i')}P_{i,i'}=P_{i,i'}$, and so forth.

Using the projectors we define the series of scalar functions related to
the resolvent
\begin{equation}
\begin{split}
g(z)\,&=P_0R(z)P_0\\
g_{i}(z)\,&=P_{i}\frac{{P_0}^C}{z-L}P_{i}\\
g_{i,i'}(z)\,&=P_{i,i'}\frac{P_{(i)}\,{P_i}^C}{z-L}P_{i,i'}\\
g_{i,i',i''}(z)\,&=
P_{i,i',i''}\frac{P_{(i,i')}\,{P_{i,i'}}^C}{z-L}P_{i,i',i''}\\
&\vdots\qquad.
\end{split}
\label{eq:g-definitions}
\end{equation}
From (\ref{eq:partitioning-theorem}) and the assumptions
(\ref{eq:assume-tree}) we have the chain of equations for these functions

\begin{equation}
\begin{split}
g(z)\,&=\frac{1}{z-\sum_{i=1}^{k}\,g_i(z)}\\
g_i(z)\,&=\frac{1}{z-\sum_{i'=1}^{k_i-1}\,g_{i,i'}(z)}\\
g_{i,i'}(z)\,&=\frac{1}{z-\sum_{i''=1}^{k_{i,i'-1}}\,g_{i,i',i''}(z)}\\
&\vdots\qquad.
\end{split}
\label{eq:g-chain}
\end{equation}
On a random tree, the degrees $k$, $k_i$, $k_{i,i'}$ are random
variables and
therefore also $g(z)$,  $g_{i}(z)$, $g_{i,i'}(z)$, etc. are random
functions of $z$. To describe their properties, we define their
generating functions (dependence on $z$ becomes implicit)
\begin{equation}
\begin{split}
G(\omega)\,&=\langle e^{-\omega\,g(z)} \rangle\\
G_1(\omega)\,&=\langle e^{-\omega\,g_i(z)} \rangle\\
G_2(\omega)\,&=\langle e^{-\omega\,g_{i,i'}(z)} \rangle\\
G_3(\omega)\,&=\langle e^{-\omega\,g_{i,i',i''}(z)} \rangle\\
&\vdots\qquad.
\end{split}
\label{eq:generating-functions-def}
\end{equation}
If the graph in question is the Erd\H{o}s-R\'enyi random graph, all
the degrees in the corresponding random tree are independent and
distributed according to the Poisson distribution
$P(k)=e^{-\mu}\,\mu^k/k!$. The average degree $\mu$ is the only free
parameter of this model.

Calculation of the generating functions 
(\ref{eq:generating-functions-def})  is facilitated by the integral representation
\begin{equation}
g(z)=\frac{1}{z-\sum_{i=1}^{k}\,g_i(z)}=\int_0^\infty
e^{-\lambda\left(z-\sum_{i=1}^{k}\,g_i(z)\right)}\, d\lambda
\end{equation}
and similarly for the other $g$'s.
Assuming for the moment that $k$ is fixed, we get, after some algebra,
the following relation between $G(\omega)$ and  $G_1(\omega)$ 
\begin{equation}
G(\omega)=1+\sqrt{\omega}\int_0^\infty\, \frac{d\lambda}{\sqrt{\lambda}}\,
I_1(2\sqrt{\omega\lambda})\,
e^{-\lambda\,z}\,\left[G_1(\lambda)\right]^k\;.
\end{equation}
Now we take into account the Poisson distribution of degrees,
which gives
\begin{equation}
G(\omega)=1+\sqrt{\omega}\int_0^\infty\, \frac{d\lambda}{\sqrt{\lambda}}\,
I_1(2\sqrt{\omega\lambda})\,
e^{-\lambda\,z+\mu\left(G_1(\lambda)-1\right)}\;.
\end{equation}
Repeating the same steps for further generating functions we get
\begin{equation}
G_1(\omega)=1+\sqrt{\omega}\int_0^\infty\, \frac{d\lambda}{\sqrt{\lambda}}\,
I_1(2\sqrt{\omega\lambda})\,
e^{-\lambda\,z+\mu\left(G_2(\lambda)-1\right)}\;.
\end{equation}
Note that the form of the relation between $G$ and $G_1$ is the same
as between $G_1$ and $G_2$ and generally between $G_m$ and $G_{m+1}$
for any $m>0$. This is due to special property of the Poisson
distribution, $kP(k)/\mu=P(k-1)$. For any other distribution this does not hold.

For infinitely large tree we suppose that the generating functions
$G_m$, $m=1,2,3,\ldots$  converge to a common limit and we can impose
the condition of stationarity $G_1(\omega)=G_2(\omega)$. Therefore, we
define a single function $\gamma(\omega)=G(\omega)-1$, for which
we have a closed equation 

\begin{equation}
\gamma(\omega)=\sqrt{\omega}\int_0^\infty\, \frac{d\lambda}{\sqrt{\lambda}}\,
I_1(2\sqrt{\omega\lambda})\,
e^{-\lambda\,z+\mu\gamma(\lambda)}\;.
\label{eq:integral-equation}
\end{equation}
It is strictly equivalent to the Equation (18) in \cite{rod_bra_88}
(the Rodgers-Bray equation),
which was obtained using the replica method. Hence we conclude that
explicit calculation showed equivalence of
replica and cavity approaches  in the case of
Erd\H{o}s-R\'enyi graph, which is just the situation in which the
Rodgers-Bray equation holds. 
Note however, that the direct computation
we used here would fail if the degree distribution was not
Poissonian.

\section{Variational problem}

The key result (\ref{eq:integral-equation}) can be reformulated in
a different way, more appropriate for approximate solution. As a first
step, we define an auxiliary function $\rho(\omega)=e^{-\omega
  z+\mu\,\gamma(\omega)}$. Instead of the single equation
(\ref{eq:integral-equation}), we can solve the pair
\begin{equation}
\begin{split}
&\gamma(\omega)=\sqrt{\omega}\int_0^\infty\, \frac{d\lambda}{\sqrt{\lambda}}\,
I_1(2\sqrt{\omega\lambda})\,\rho(\lambda)\\
&\rho(\omega)=e^{-\omega
  z+\mu\,\gamma(\omega)}\;.
\label{eq:pair-equations-gamma-rho}
\end{split}
\end{equation}
Direct solution of (\ref{eq:pair-equations-gamma-rho}) is as difficult
as solving (\ref{eq:integral-equation}). However, we
can find a functional, whose stationary point is just defined by
equations  (\ref{eq:pair-equations-gamma-rho}). We can check
explicitly that such functional is
\begin{equation}
\begin{split}
\mathcal{F}&[\gamma,\rho]=
-\int_0^\infty\frac{d\omega}{\omega}\,\gamma(\omega)\rho(\omega)+\\
&+\frac{1}{2}\int_0^\infty\frac{d\omega}{\sqrt{\omega}}
\int_0^\infty\frac{d\lambda}{\sqrt{\lambda}}
\,I_1(2\sqrt{\omega\lambda})\rho(\omega)\rho(\lambda)+\\
&+\frac{1}{\mu}\int_0^\infty\frac{d\omega}{\omega}e^{-\omega
  z+\mu\,\gamma(\omega)}\;.
\label{eq:functional}
\end{split}
\end{equation}
Note that we derived, within the cavity approach, a result which is
analogous to the functional obtained in \cite{kuhn_08} using the
replica trick. 

The variational formulation of the problem is useful as a generator of
approximations. In \cite{sem_cug_02} a variational ansatz was used to
derive the density of states in effective-medium approximation (EMA). Let us
see now how it is obtained in our cavity procedure.
If we take the exponential ansatz for the auxiliary function
$\rho(\omega)$, namely  
\begin{equation}
\rho(\omega)=e^{-\sigma\,\omega}
\label{eq:ansatz-ema}
\end{equation}
all integrals in (\ref{eq:functional}) can be performed explicitly and
we can extremalize the functional with respect to $\sigma$ and
$\gamma(\omega)$. This way we find the cubic equation
\begin{equation}
\sigma^3-z\,\sigma^2+(\mu-1)\,\sigma+z=0\;.
\label{eq:equation-for-sigma-ema}
\end{equation}
It is identical to the equation (23) in \cite{sem_cug_02} obtained by
the replica method. The solution can be obtained analytically and the
density of states is extracted using the formula
\begin{equation}
\mathcal{D}(z)=\lim_{\epsilon\to0^+}\mathrm{Im}\frac{1}{\pi\,\sigma(z-i\epsilon)}\;.
\end{equation}

We can further improve the calculation by the following trick, which
we shall refer as ``single-shell approximation'' within this paper. We may
formally write the pair of equations (\ref{eq:pair-equations-gamma-rho})
as a set of four equations
\begin{equation}
\begin{split}
&\gamma(\omega)=\sqrt{\omega}\int_0^\infty\, \frac{d\lambda}{\sqrt{\lambda}}\,
I_1(2\sqrt{\omega\lambda})\,\rho(\lambda)\\
&\rho(\omega)=e^{-\omega
  z+\mu\,\overline{\gamma}(\omega)}\\
&\overline{\gamma}(\omega)
=\sqrt{\omega}\int_0^\infty\, \frac{d\lambda}{\sqrt{\lambda}}\,
I_1(2\sqrt{\omega\lambda})\,\overline{\rho}(\lambda)\\
&\overline{\rho}(\omega)=e^{-\omega
  z+\mu\,\gamma(\omega)}\;\qquad\qquad.
\label{eq:four-equations-gamma-rho-AB}
\end{split}
\end{equation}
These equations can be obtained as a condition of stationarity for the
functional

\begin{equation}
\begin{split}
\mathcal{F}_1[\gamma,\rho,&\overline{\gamma},\overline{\rho}]=
-\int_0^\infty\frac{d\omega}{\omega}\Big(
\gamma(\omega)\overline{\rho}(\omega)+\overline{\gamma}(\omega)\rho(\omega)\Big)\\
&+\int_0^\infty\frac{d\omega}{\sqrt{\omega}}
\int_0^\infty\frac{d\lambda}{\sqrt{\lambda}}
\,I_1(2\sqrt{\omega\lambda})\rho(\omega)\overline{\rho}(\lambda)+\\
&+\frac{1}{\mu}\int_0^\infty\frac{d\omega}{\omega}e^{-\omega  z}
\Big(e^{\mu\,\gamma(\omega)}+e^{\mu\,\overline{\gamma}(\omega)}\Big)
\end{split}
\label{eq:functional-two}
\end{equation}

If the equations (\ref{eq:four-equations-gamma-rho-AB}) were solved exactly, we would have
$\gamma(\omega)=\overline{\gamma}(\omega)$ and
$\rho(\omega)=\overline{\rho}(\omega)$. The same would hold also in the
case of 
the effective medium approximation, which amounts taking the ansatz
$\rho(\omega)=\overline{\rho}(\omega)=e^{-\sigma\,\omega}$, so seemingly
the set (\ref{eq:four-equations-gamma-rho-AB}) does not bring any
advantage over (\ref{eq:pair-equations-gamma-rho}). However, relaxing
the condition $\rho(\omega)=\overline{\rho}(\omega)$ we can get an
improvement in an approximate solution. Indeed, we can take the ansatz
\begin{equation}
\rho(\omega)=e^{-\sigma\,\omega}
\end{equation}
as in EMA, but allow $\overline{\rho}(\omega)$
adjust itself freely so that $\mathcal{F}_1$ is stationary. This way
we introduce an error, because $\rho(\omega)\ne\overline{\rho}(\omega)$ and 
$\gamma(\omega)\ne\overline{\gamma}(\omega)$, but we gain better
approximation for the density of states.

After some algebra, we get the following equation for the quantity
$\tau=z\,\sigma$

\begin{equation}
z^2=
\mu+\tau+e^{-\mu}\sum_{l=1}^\infty
\frac{\mu^l}{(l-1)!}\,\frac{l}{\tau-l}\;.
\label{eq:equation-for-tau}
\end{equation}
The fact that the equation depends
on $z^2$ means that the spectrum is symmetric with respect to the
point $z=0$. For a general $z$ on the real axis the equation
(\ref{eq:equation-for-tau}) can be easily solved numerically. We find 
that there are at most two roots with non-zero imaginary parts
(complex conjugate to each other). Those values of $z$ for which all
solutions are real correspond to gaps in the spectrum. General picture
is that there is a very narrow gap around $z=0$, separating two halves
of a wide band, containing most of the eigenvalues. We can call this
band (not quite precisely) as ``bulk'' of the density of states.

In the middle of the bulk, there is a $\delta$-function
contribution just at $z=0$, whose weight can be
found exactly and is equal to $\mathrm{e}^{-\mu}$. On both sides of
the bulk, there are series of small side bands separated by
gaps.  The density of states has therefore the form
\begin{equation}
\mathcal{D}(z)=\mathrm{e}^{-\mu}\delta(z)+\mathcal{D}_c(z)
\end{equation}
where $\mathcal{D}_c(z)$ is a continuous function.
 The interpretation of the $\delta$-function is
straightforward. It corresponds to single isolated vertices, whose
fraction is just equal to $\mathrm{e}^{-\mu}$ and they all contribute
with the same eigenvalue $0$.  

Some analytical information on the continuous part $\mathcal{D}_c(z)$
can be found from approximate solution of the equation
(\ref{eq:equation-for-tau}). For $\mathrm{e}^{-\mu}\ll 1$ we can find approximately the edge of the
gap around $z=0$. We get
\begin{equation}
\mathcal{D}_c(z)\simeq\frac{1}{2\pi\,z}\sqrt{4\psi(\mu)\,z^2-\mathrm{e}^{-2\mu}} 
\end{equation}
where we denoted 
\begin{equation}
\psi(\mu)=\mathrm{e}^{-\mu}
\,\sum_{l=1}^\infty\frac{\mu^l}{l!\,l}=\mu\,\mathrm{e}^{-\mu}
\,{}_2\! F_2(1,1;2,2;\mu) \;.
\end{equation}
%
%
%
We can see that the gap edge is at $z_0=\frac{1}{2}\mathrm{e}^{-\mu}/\sqrt{\psi(\mu)}$.

For the tails, we can calculate analytically the side bands in an
 approximation which becomes exact for 
 $|z|\to\infty$. The computation goes as follows. 
Each of the side bands can be identified with one term in the infinite
sum over $l$ in (\ref{eq:equation-for-tau}). The tails of the spectrum
corresponding to large $|z|$  are identified with large $l$. 
In the crudest approximation,
the solution is $\tau\simeq l$. Therefore, we introduce a new variable
$\eta$ by $\tau=l+\eta$. So,  (\ref{eq:equation-for-tau}) assumes the
form
\begin{equation}
\begin{split}
z^2=\,&\mu+l+\eta+e^{-\mu}
\frac{\mu^l}{(l-1)!}\,\frac{l}{\eta}+\\
&+e^{-\mu}\sum_{\substack{l'=1\\(l'\ne l)}}^\infty
\frac{\mu^{l'}}{(l'-1)!}\,\frac{l'}{l-l'-\eta}
\end{split}
\end{equation}
For large $l$ we can expand the infinite series in powers of $\eta$
and keep only the lowest terms, so
\begin{equation}
\begin{split}
z^2=\,&\mu+\Delta_l(\mu)+l+\big(1-\Gamma_l(\mu)\big)\,\eta+\\
&+e^{-\mu}
\frac{\mu^l}{(l-1)!}\,\frac{l}{\eta}+O(\eta^2)
\end{split}
\label{eq:equation-expansion-in-eta}
\end{equation}
where
\begin{equation}
\begin{split}
&\Delta_l(\mu)=e^{-\mu}\sum_{\substack{l'=1\\(l'\ne l)}}^\infty
\frac{\mu^{l'}}{(l'-1)!}\,\frac{l'}{l-l'}\\
&\Gamma_l(\mu)=e^{-\mu}\sum_{\substack{l'=1\\(l'\ne l)}}^\infty
\frac{\mu^{l'}}{(l'-1)!}\,\frac{l'}{(l-l')^2}\;.
\end{split}
\end{equation}

So, for each $l$, large enough, we have two  ``bubbles'' of non-zero density of
states. The two bubbles are symmetric to each other with respect to
the origin. The ``bubbles'' are separated by gaps, so each ``bubble'' has
well defined lower and upper edges, $z_{l-}$ and $z_{l+}$, respectively.
For large $l$ the
approximate form of the ``bubble'' is given by the solution of
a quadratic equation in $\eta$, so
\begin{equation}
\begin{split}
\mathcal{D}_{l}(z)\,&\simeq\frac{|z|}{\pi}\;\Bigg[(1-\Gamma_{l}(\mu))
\frac{\mathrm{e}^{-\mu}\,l\,\mu^{l}}{(l-1)!}-\\
&-\bigg(\frac{z^2-\mu-l-\Delta_{l}(\mu)}{2}\bigg)^2\Bigg]^{1/2}\times\\
&\times\Bigg[
\frac{\mathrm{e}^{-\mu}\,l\,\mu^{l}}{(l-1)!}+(z^2-\mu-l)l+\\
&+(1-\Gamma_{l}(\mu))(l)^2
\Bigg]^{-1}\;.
\end{split}
\end{equation}
The width of the bubble  $z_{l+}-z_{l-}$ approaches zero for $l\to
\infty$. This justifies considering $\eta$ a small parameter in the
expansion (\ref{eq:equation-expansion-in-eta}). For large $l$ the
``bubbles'' have a semi-circle shape and their weight is
\begin{equation}
W_{l}=
\int_{z_{l-}}^{z_{l+}}
\mathcal{D}_{l}(z)\;dz
\simeq
\frac{1}{2}\, e^{-\mu}\,\frac{\mu^{l}}{l!}\;.
\end{equation}
We recognize the Poisson distribution with mean $\mu$. This reflects
the Poisson distribution of degrees of the random graph. The factor
$1/2$ stems from the fact that we have two bubbles for each $l$. 
The center of the bubble corresponding to $l$ is at
$z_{l}=\sqrt{l+\mu+\Delta_{l}(\mu)}$, thus the distance between
  two successive bubbles is $\Delta z_{l}\simeq
  (4z_{l})^{-1/2}$. Hence we deduce the approximate density of
  states in the tails, for $|z|\to\infty$
\begin{equation}
\mathcal{D}_\mathrm{tail}(z)\simeq
\frac{\mathrm{e}^{-\mu}\,|z|\,\mu^{z^2}}{\Gamma(z^2-1)}
\simeq
\frac{\mathrm{e}^{-\mu}}{\sqrt{2\pi}}
\,\bigg(\frac{e\mu}{z^2}\bigg)^{z^2}\;.
\end{equation}
This is the shape of the Lifschitz tail, which was already obtained 
 by \cite{rod_bra_88} and \cite{sem_cug_02}.

\begin{figure}[t]
\includegraphics[scale=0.85]{%
\slaninafigdir/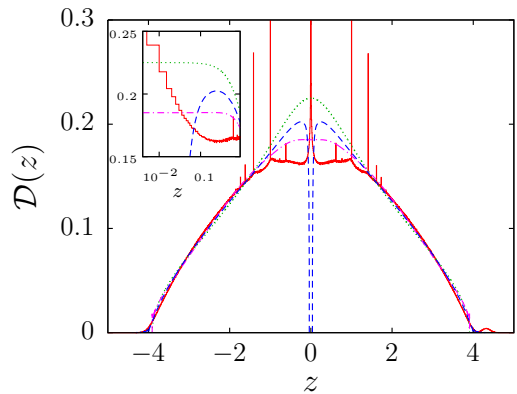}
\caption{
Density of states for the adjacency matrix of Erd\H os-R\'enyi
graph, with average degree $\mu=3$.
 Full line shows the result of numerical diagonalization of
matrix of size $N=1000$, averaged over $75000$ random
realizations. The dotted line is the result of effective medium
approximation, the dot-dashed line id the single-defect approximation, and 
the dashed line is the single-shell approximation. In the inset,
detail of the density of states around the center of the band, plotted
in semi-logarithmic scale.  
}
\label{fig:dos-3}
\end{figure}

To assess the quality of the approximations used, we compare the
 results arising from EMA (Eq. (\ref{eq:equation-for-sigma-ema})),
 from the single-defect \cite{sem_cug_02,rod_cas_kuh_tak_08}, and
 single-shell (Eq. (\ref{eq:equation-for-tau})) approximations with
 average density of states obtained by numerical diagonalization of
 sample matrices. In Fig. \ref{fig:dos-3} we can see the spectrum for
 $\mu=3$ and matrices of size
 $N=1000$ averaged over $75000$ realizations. We can clearly identify
 the delta-peaks, as well as the complicated shape of the continuous
 part of the spectrum near the center of the bulk. Interestingly, both
 EMA and single-shell approximations are very good if we are neither
 close to the center nor at the tails of the spectrum. Close to the
 center, the shape of the density of states is rather complex, as
 shown in the inset in Fig. \ref{fig:dos-3}. There is a shallow
 depression, followed by a divergence at $z=0$. The form of the
 singularity at $z=0$ seems to
 be close to a logarithmic divergence, although the data do not
 provide a decisive evidence. Neither of the three approximations
 reproduces this singularity. EMA and SDA are constant around $z=0$,
 while the single-shell approximation exaggerates the depression
 around $z=0$ to such extent that a spurious gap is created. This is
 an artifact of the approximation.
However,
 the delta-peak at the origin is, correctly, present in the
 single-shell approximation.

\begin{figure}[t]
\includegraphics[scale=0.85]{%
\slaninafigdir/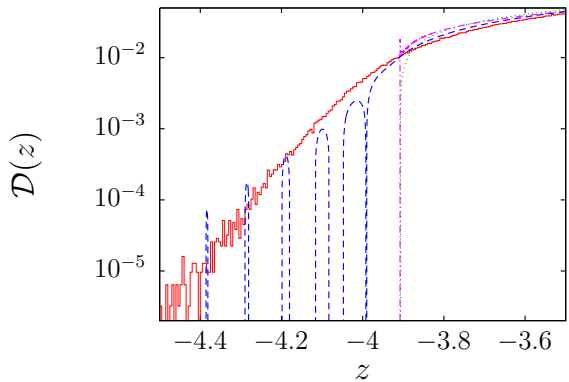}
\caption{
The detail of the left tail of the density of states shown in
Fig. \ref{fig:dos-3}.
The full line shows the result of numerical diagonalization, 
the dotted line is the result of effective medium
approximation, the dot-dashed line id the single-defect approximation, and 
the dashed line is the single-shell approximation.
}
\label{fig:dos-3-tail}
\end{figure}

Similar comparison was done also at the tail of the density of
states. We can see in Fig. \ref{fig:dos-3-tail} a detail of the same
data as shown in Fig.  \ref{fig:dos-3}. Note that, for any finite $N$,
the density of states  is not
mirror-symmetric with respect of the line $z=0$, because the average
value of the off-diagonal elements of the matrix $L$ is strictly
positive. Only in the limit $N\to\infty$ the spectrum becomes
symmetric. The single largest eigenvalue is split off the rest of the
spectrum \cite{fur_kom_81} and the small bump in the positive tail
corresponds to this effect. In the limit $N\to\infty$ this bump would
vanish, as the weight of the single largest eigenvalue becomes
negligible compared to the rest of the spectrum.

As shown in  Fig. \ref{fig:dos-3-tail}, we can see that the
single-shell approximation is superior to both EMA and SDA in the tail
region, from two aspects. First, the spurious band edge of EMA and
SDA is shifted towards larger $|z|$, so that the interval in which
$D(z)$ is well reproduced is wider. Second, the single-shell approximation
displays non-zero density of states also in some regions of the
Lifschitz tails, although, instead of exhibiting a smooth behavior everywhere, the
density of states is concentrated in ``bubbles''. The gaps separating
the ``bubbles'' are again artifacts of the approximation, to the same
extent as the sharp band edge is an artifact of EMA and SDA. On the other
hand, it is an important improvement over SDA \cite{sem_cug_02}. The
delta-peaks of SDA are widened into continuous bands in our
approach. In fact, this is to be expected, because the single-shell
approximation can be rightly interpreted as a self-consistent version
of SDA. Therefore, it should be better than SDA in principle, although
this a priori judgement may turn incorrect
in practice, as  the single-shell approximation  is better than SDA 
sometimes (in the tail) but worse elsewhere
(around $z=0$).  

Finally, let us note that similar ``bubbles'' at the tails were also
seen in approximations derived using replica method by \cite{dean_02} for the Laplacian of a
random graph and by \cite{nag_tan_07} for  sparse covariance matrices.

\section{Covariance matrices}

Another application of the method presented here is investigation of
sparse covariance matrices. They can be considered as arising from a
bipartite graph where edges connect vertices from the set
$A$ with vertices from the set $B$. We denote the size of the sets
$N_A$ and $N_B$, respectively. In the thermodynamic limit,
$N_A\to\infty$,  $N_B\to\infty$, we fix the 
ratio $\alpha=N_A/N_B$ constant. In the bipartite analog of
Erd\H{o}s-R\'enyi random graph, the degrees if vertices in $A$ and $B$
follow Poisson distributions with average degree $\mu_A$ and $\mu_B$,
respectively, where $\mu_B/\mu_A=\alpha$. The problem has a long
history, starting with the work of Mar\v{c}enko and Pastur
\cite{mar_pas_67}
  and was investigated recently
by replica method in  \cite{nag_tan_07}.

The adjacency matrix of the bipartite graph has the form
\begin{equation}
L=\Bigg(
\begin{array}{ll}
0&M^T\\
M&0
\end{array}
\Bigg)
\label{eq:matrix-on-bipartite}
\end{equation}
where the first block of indices corresponds to set $A$, second block
to set $B$. We define the contraction, or covariance, matrix
$C_A=M^TM$, which acts solely in the set $A$ (and similarly
$C_B=MM^T$, which acts solely in the set $B$. The spectra of the
matrices $L$, $C_A$ and $C_B$  are closely related. We define
$\mathcal{D}_A(z)=\lim_{\epsilon\to 0^+}\mathrm{Im}\sum_{i\in
  A}[(z-i\epsilon-L)^{-1}]_{ii}/(N_A\,\pi)
$
the partial density of states of $L$ restricted to the set $A$ and
$\mathcal{D}_{CA}(z)=\lim_{\epsilon\to 0^+}\mathrm{Im}\sum_{i\in
  A}[(z-i\epsilon-C_A)^{-1}]_{ii}/(N_A\,\pi)$ the density of states of the
correlation matrix $C_A$. It can be easily shown that
\begin{equation}
\mathcal{D}_{CA}(z)=\frac{1}{\sqrt{z}}\mathcal{D}_A(\sqrt{z})\;.
\label{eq:relation-corrmatrix-matrix}
\end{equation}
This relation remains in force also after averaging over the randomness
in the matrix $M$. Therefore, to calculate the average density of
states of the covariance matrix $C_A$ it is enough to investigate the
matrix element $\langle[(z-L)^{-1}]_{ii}\rangle$ for any $i\in A$. To
this end, we define the generating functions
\begin{equation}
\begin{split}
&\gamma_A=\langle e^{-\omega[(z-L)^{-1}]_{ii}}\rangle-1\text{ for }i\in A\\
&\gamma_B=\langle e^{-\omega[(z-L)^{-1}]_{jj}}\rangle-1\text{ for }j\in B\;.
\end{split}
\end{equation}
Further procedure follows closely that of the previous
section. Finally, we get a set of four coupled equations, very similar
to the set we encountered in the single-shell approximation
\begin{equation}
\begin{split}
&\gamma_A(\omega)=\sqrt{\omega}\int_0^\infty\, \frac{d\lambda}{\sqrt{\lambda}}\,
I_1(2\sqrt{\omega\lambda})\,\rho_B(\lambda)\\
&\rho_B(\omega)=e^{-\omega
  z+\mu_A\, \gamma_B (\omega)}\\
&\gamma_B(\omega)
=\sqrt{\omega}\int_0^\infty\, \frac{d\lambda}{\sqrt{\lambda}}\,
I_1(2\sqrt{\omega\lambda})\,\rho_A(\lambda)\\
&\rho_A(\omega)=e^{-\omega
  z+\mu_B\,\gamma_A(\omega)}\;\qquad\qquad.
\end{split}
\label{eq:four-equations-gamma-rho-bipartite}
\end{equation}
We can easily check that the solution of these equations makes
the following functional stationary
\begin{equation}
\begin{split}
&\mathcal{F}_{AB}[\gamma_A,\rho_A,{\gamma_B},{\rho_B}]=\\
&-\int_0^\infty\frac{d\omega}{\omega}\,\Big(
\gamma_A(\omega){\rho_A}(\omega)+{\gamma_B}(\omega)\rho_B(\omega)\Big)\\
&+\int_0^\infty\frac{d\omega}{\sqrt{\omega}}
\int_0^\infty\frac{d\lambda}{\sqrt{\lambda}}
\;I_1(2\sqrt{\omega\lambda})\rho_A(\omega){\rho_B}(\lambda)+\\
&+\int_0^\infty\frac{d\omega}{\omega}e^{-\omega  z}
\Big(\frac{1}{\mu_A}\,e^{\mu_A\,\gamma_B(\omega)}+\frac{1}{\mu_B}\,e^{\mu_B\,{\gamma_A}(\omega)}\Big)\;.
\end{split}
\label{eq:functional-bipartite}
\end{equation}
For an approximate solution of the equations
(\ref{eq:four-equations-gamma-rho-bipartite}) we use again a
variational ansatz. In analogy with EMA, we assume the following form
\begin{equation}
\begin{split}
&\rho_A(\omega)=e^{-\sigma_A\,\omega}\\
&\rho_B(\omega)=e^{-\sigma_B\,\omega}\;.
\end{split}
\label{eq:ansatz-bipartite-ema}
\end{equation}
The insertion of (\ref{eq:ansatz-bipartite-ema}) in
(\ref{eq:functional-bipartite}) produces finally two uncoupled cubic
equations for $\sigma_A$ and $\sigma_B$. The equation relevant for us is
\begin{equation}
\begin{split}
z\,\sigma_B^3 +\big((1-\alpha)\mu_A+\alpha-1-z^2\big)
\,\sigma_B^2 +&\\
+\big(\mu_A\,\alpha+1-2\alpha\big)
\,z\,\sigma_B 
+&z^2\,\alpha=0
\end{split}
\label{eq:cubic-on-bipartite}
\end{equation}
where we used $\alpha=\mu_B/\mu_A$. The average density of states for
the covariance matrix $C_A$ is found considering the first equation of
(\ref{eq:four-equations-gamma-rho-bipartite}) and the relation
(\ref{eq:relation-corrmatrix-matrix}), thus 
\begin{equation}
\mathcal{D}_{CA}(z)=\frac{1}{\pi\sqrt{z}}
\lim_{\epsilon\to  0+}\mathrm{Im}
\frac{1}{\sigma_B(\sqrt{z}-i\epsilon)}\;.
\label{eq:dos-corrmatrix}
\end{equation}
The solution can be obtained analytically, but we shall not show the
formula here. However, we can check that in the limit $\mu_A\to\infty$
with $\alpha$ and $\zeta=z/\mu_A$ fixed we get
\begin{equation}
\begin{split}
\mathcal{D}_{CA}(\zeta)\,&=
\frac{1}{2\pi\alpha\,\zeta}\times\\
&\times\sqrt{\Big((1+\sqrt{\alpha})^2-\zeta\Big)\Big(\zeta-(1-\sqrt{\alpha})^2\Big)} 
\end{split}
\label{eq:marcenko-pastur}
\end{equation}
which is the Mar\v{c}enko-Pastur (MP) density of states \cite{mar_pas_67}.

\begin{figure}[t]
\includegraphics[scale=0.85]{%
\slaninafigdir/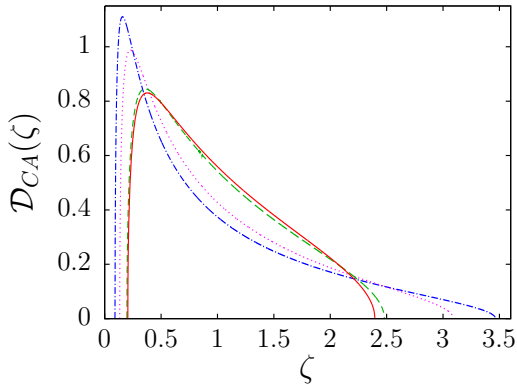}
\caption{
Density of states for the correlation matrix based on sparse adjacency
matrix, for $\alpha=0.3$. The average degree is $\mu_A=3$ (dash-dotted
line), $5$ (dotted line), and $50$ (dashed line). The full line is the
Mar\v{c}enko-Pastur density (\ref{eq:marcenko-pastur}), i. e. the
limit $\mu_A\to\infty$.}
\label{fig:corrmat-03-all}
\end{figure}

In Fig. \ref{fig:corrmat-03-all} we show the density of states as
function of $\zeta=z/\mu_A$ for several values of $\mu_A$, as found by
solution of (\ref{eq:cubic-on-bipartite}). We can see
that the approach to MP density is rather slow. We found that the
difference can be considered small only at about $\mu_A\simeq 50$.

\section{Conclusions}

We considered a random graph of large size $N\to\infty$ of two
types. First, a ``classical'' Erd\H{o}s-R\'enyi graph, and second, a
random bipartite graph. We calculated density of eigenvalues of
adjacency matrices of these graphs. In the case of the bipartite
graph, the final result was the density of states of the covariance
matrix, defined by a contraction of the adjacency matrix.

Our contribution to the problem of spectra of sparse random matrices
consists in showing that the cavity approach, i. e. approximation of
the random graph by a random tree, is exactly equivalent to the
calculation by replica method in the thermodynamic limit. Furthermore,
we demonstrated how the cavity calculation can be formulated as a
variational problem, similar but substantially simpler than the
variational formulation arising from the replica method. At minimum,
we do not need to consider the possibility of replica-breaking
solutions, which are known to exist and contribute to the finite-size
corrections \cite{kam_mez_99}. We can interpret it also in the
following manner. Since we are working directly with
infinite-size system, $N=\infty$, the physics behind the
replica-breaking states has no effect. 

The variational formulation introduced here is a very practical
starting point for approximations. The exponential ansatz leads to
results identical to the effective-medium approximation studied
earlier \cite{sem_cug_02}. However, using our variational scheme, the
approximation can be easily improved by what we call ``single-shell
approximation''. It produces the Lifschitz tail in the density of
states in the form of a series of ``bubbles''. We are able to
calculate  the weight and
distance of the bubbles. Hence we arrive at average density of states in the
tail, which is identical to the old result of Rodgers and Bray
\cite{rod_bra_88}. 
Furthermore, we applied the method also to the spectra of sparse
covariance matrices, where we easily derived a formula generalizing the
Mar\v{c}enko-Pastur density of states. 

The variational formulation introduced here can be used not only as a
generator of approximations, but also as a basis of numerical
methods. Indeed, there is no principal obstacle for numerical
extremalization of the functional of two variables. This contrasts
with the variational methods based on replica trick, where the replica
limit $n\to 0$, involving analytic continuation,  must be done after
extremalization, whch makes the method numerically unfeasible.

We believe that the method can be applied also for other types of
random graphs.
We must, however, admit a serious limitation of our method, which is
the Poisson distribution of degrees of the graph. Therefore, it is,
for example, not applicable directly for graphs with power-law degree
distribution. We believe that the roots of this limitation lie quite
deep. For example, to our best knowledge, there is no replica
calculation available for random graphs defined by their degree
sequence only. And, on the other side, there are no results from
cavity method for those random graphs with power-law degree
distribution, for which replica calculations do exist, like those of
Ref.\cite{rod_aus_kah_kim_05}. The point is, that for
Erd\H{o}s-R\'enyi graph, it is well established that the local
topology is isomorphic to a random tree. For a graph with general
degree sequence, non obeying Poisson statistics, this may or may not
be true. The question of equivalence or not of replica and cavity
methods is intimately related to the question of local isomorphism to
a tree, which is rather complex and not solved in general. 
Hence, a  successful treatment of such cases by
both replica and cavity method in parallel, would require, very probably,  completely
novel ideas.

\begin{acknowledgments}
I gladly acknowledge inspiring discussions with J. Ma\v{s}ek. 
This work was carried out within the project AV0Z10100520 of the Academy of 
Sciences of the Czech Republic and was 
supported by the M\v{S}MT of the Czech Republic, grant no. 
OC09078.

\end{acknowledgments}
\end{document}